%% file: main_wif_mfi2020_final_conf_format.tex
\documentclass[letterpaper, 10 pt, conference]{ieeeconf}  

\IEEEoverridecommandlockouts                              

\title{\LARGE \bf
Weighted Information Filtering, Smoothing, and Out-of-Sequence Measurement Processing}

\author{Yaron Shulami$^{1}$ and Daniel Sigalov$^{2}$
\thanks{$^{1}$Yaron Shulami is with Rafael -- Advances Defense Systems, Haifa, Israel
        {\tt\small yaronsho@rafael.co.il}}%
\thanks{$^{2}$Daniel Sigalov is with Rafael -- Advances Defense Systems, Haifa, Israel
        {\tt\small danielsi@rafael.co.il}}%
}

\input{packages}

\input{commands}

\usepackage[font=bf]{subfig}

\DeclareGraphicsExtensions{.eps}
\graphicspath{{../figures/}}

\newtheorem{theo}{Theorem}

\newtheorem{lemm}{Lemma}

\begin{document}

\maketitle
\thispagestyle{empty}
\pagestyle{empty}

\begin{abstract}
We consider the problem of state estimation in dynamical systems and propose a different mechanism for handling unmodeled system uncertainties. Instead of injecting random process noise, we assign different weights to measurements so that more recent measurements are assigned more weight. A specific choice of exponentially decaying weight function results in an algorithm with essentially the same recursive structure as the Kalman filter. It differs, however, in the manner in which old and new data are combined. While in the classical KF, the uncertainty associated with the previous estimate is inflated by adding the process noise covariance, in the present case, the uncertainty inflation is done by multiplying the previous covariance matrix by an exponential factor. This difference allows us to solve a larger variety of problems using essentially the same algorithm. We thus propose a unified and optimal, in the least-squares sense, method for filtering, prediction, smoothing and general out-of-sequence updates. All of these tasks require different Kalman-like algorithms when addressed in the classical manner.
\end{abstract}

\input{paper_text}

\addtolength{\textheight}{-12cm}   


\section*{ACKNOWLEDGMENT}
The authors would like to thank Ms. Shirley Rosenfeld for improving the grammar of the paper and correcting numerous topographical errors.
The constructive comments of the anonymous reviewers are highly appreciated.


\end{document}

%% file: packages.tex
\usepackage[english]{babel}
\usepackage{url}
\usepackage[cp1255]{inputenc}
\usepackage{amsfonts}
\usepackage{amssymb}
\usepackage{amsmath}
\usepackage{verbatim}
\usepackage{graphics}
\usepackage{graphicx}
\usepackage{bbm}
\usepackage{cancel}
\usepackage{fancyhdr}

\let\ORIGRightarrow=\Rightarrow

\let\Rightarrow=\ORIGRightarrow

\usepackage[usenames]{color}

\usepackage{algorithmic}
\usepackage{algorithm}

%% file: commands.tex
\newcommand{\EE}[1]{\mathbb{E}\left[#1\right]}

\newcommand{\beq}{\begin{equation}}
\newcommand{\eeq}{\end{equation}}

\newcommand{\beqa}{\begin{eqnarray}}
\newcommand{\eeqa}{\end{eqnarray}}

\newcommand{\non}{\nonumber}

\newcommand{\size}[1]{\left|#1\right|}

\newcommand{\set}[1]{\left\{#1\right\}}

\newcommand{\paren}[1]{\left(#1\right)}

\newcommand{\gaus}[2]{{\mathcal{N}}\paren{#1,#2}}


%% file: paper_text.tex
\renewcommand{\algorithmicrequire}{\textbf{Input:}}
\renewcommand{\algorithmicensure}{\textbf{Output:}}

\section{Introduction}\label{section:intro}
\input{intro_mfi}

\section{Problem Setup}\label{section:problem}
\input{problem_mfi}

\section{Derivation of the Estimator}\label{section:derivation}
\subsection{Batch-Form Estimator}\label{section:batch:derivation}
We begin with the derivation of the optimal estimator, in batch form, for the quadratic cost~\eqref{Eq:problem:WLS}. For clarity of the exposition we consider~\eqref{Eq:problem:measurement2} as the measurement model. The first result, which is a straightforward generalization of the classical LS solution, is summarized in the following two Theorems.
\begin{theo}\label{theorem:batch}
Let $\set{y_\ell,\,\ell\in\mathcal{ L}}$ be a general set of measurements defined in~\eqref{Eq:problem:measurement2}. The optimal solution of the problem~\eqref{Eq:problem:WLS} subject to the constraint~\eqref{Eq:problem:state} is
\begin{align}\non
\hat{x}_k
&=\paren{\sum_{\ell\in\mathcal{L}}A_{\ell,k}^TH_\ell^TW_{\ell,k}R_\ell^{-1}W_{\ell,k}^TH_\ell A_{\ell,k}}^{-1}\\\label{Eq:batch:estimator}
&\qquad\times\sum_{\ell\in\mathcal{L}}A_{\ell,k}^TH_\ell^TW_{\ell,k}R_\ell^{-1}W_{\ell,k}^T(y_\ell-H_\ell C_{\ell,k}u_{\ell,k}).
\end{align}
\end{theo}

\begin{theo}\label{theorem:covariance}
The covariance of the estimator~\eqref{Eq:batch:estimator} is given by
\begin{align}\non
P_k&\triangleq\EE{(\hat{x}_k-\EE{\hat{x}_k})(\hat{x}_k-\EE{\hat{x}_k})^T}\\\label{Eq:batch:covariance}
&=\paren{\sum_{\ell\in\mathcal{L}}A_{\ell,k}^TH_\ell^TW_{\ell,k}R_\ell^{-1}W_{\ell}^TH_\ell A_{\ell,k}}^{-1}.
\end{align}
\end{theo}
Equation~\eqref{theorem:covariance} allows us to rewrite~\eqref{Eq:batch:estimator} in the following compact form
\begin{align}\label{Eq:batch:estimator:compact}
\hat{x}_k=P_k\sum_{\ell\in\mathcal{L}}A_{\ell,k}^TH_\ell^TW_{\ell,k}R_\ell^{-1}W_{\ell,k}^T(y_\ell-H_\ell C_{\ell,k}u_{\ell,k}).
\end{align}

For notational convenience and without loss of generality, in the sequel, we set $u_{\ell,k}=0$ for all $k\in\mathbb{N}$ and $\ell\in\mathcal{L}$.

\subsection{Recursive Estimator}\label{section:derivation:recursive}
We now consider a narrower family of the weighting matrices $W_{\ell,k}$ that depend explicitly both on the time of the currently considered measurement $y_\ell$, and on the time of the state to be estimated $x_k$, thus accounting for the relevance of $y_\ell$ on the estimation of $x_k$. As we show in the sequel, this family allows recursive computation of the estimator~\eqref{Eq:batch:estimator} and the corresponding covariance~\eqref{Eq:batch:covariance}.

Recall the considered measurement model~\eqref{Eq:problem:measurement2}. The weight assigned to $y_\ell$ reduces as the difference ${t_k-t_\ell}$ increases. In this work we take $W_{\ell,k}$ to be a scalar-valued function with the following properties: %
\begin{subequations}\label{Eq:recursive:W:properties}
\begin{align}\label{Eq:recursive:W:positive}
W_{\ell,k}&\geq0, \forall\ \ell, k\\\label{Eq:recursive:W:equal}
W_{k,k}&=1\\\label{Eq:recursive:W:telescopic}
W_{\ell,k}&=W_{\ell,m}W_{m,k}.
\end{align}
\end{subequations}
The simplest, trivial example of a function satisfying the properties~\eqref{Eq:recursive:W:positive}-\eqref{Eq:recursive:W:telescopic} is $W_{\ell,k}\equiv1$. A non trivial example of such a function, to be used in the sequel, is
\begin{align}\label{Eq:recursive:W}
W_{\ell,k} = e^{-{(t_k-t_\ell)}/2\tau},\quad\tau>0.
\end{align}
Informally, the parameter $\tau$ may be viewed as a ``weight decay factor'' or as a ``decorrelation interval'' so that the amount of information about the state, stored in the measurements, reduces as the temporal distance between these measurements and the state increases. The existence and utility of additional forms of $W_{\ell,k}$ will be considered in a future work.

The three main results of this paper are stated next.
\begin{theo}\label{theorem:recursive}
Let $\mathcal{L}=\set{1,2,\ldots,k}$. The estimator~\eqref{Eq:batch:estimator} with the weighting matrix $W_{\ell,k}$ satisfying~\eqref{Eq:recursive:W:properties} may be computed recursively as follows
\begin{align}\label{Eq:recursive:estimator}
\hat{x}_k
&=P_k\big(W_{k-1,k}^2A_{k-1,k}^TP_{k-1}^{-1}\hat{x}_{k-1}+H_k^TR_k^{-1}y_k\big),
\end{align}
where the estimator covariance, $P_k$, is given by
\begin{align}\label{Eq:recursive:covariance}
P_k^{-1}={W_{k-1,k}^2A_{k-1,k}^TP_{k-1}^{-1}A_{k-1,k}+H_k^TR_k^{-1}H_k}.
\end{align}
\end{theo}

The setup of Theorem~\ref{theorem:recursive} implies that the measurements $\set{y_\ell,\ell\in\mathcal{L}}$ arrive sequentially, such that $y_k$ is the measurement arriving at time $t_k$ thus being an observation of $x_k$ according to~\eqref{Eq:problem:measurement}. We show next that this does not have to be the case. To this end, consider a more general set of indices
\begin{align}\label{Eq:indices:new}
\mathcal{L}=\set{\ell_1,\ell_2,\ldots,\ell_k},
\end{align}
such that the measurement $y_{\ell_k}$ arrives at time $t_k$, but corresponds to time $t_{\ell_k}$. Clearly, if $\ell_k=k$ for all $\ell_k\in\mathcal{L}$, the present setup reduces to the one discussed in Theorem~\ref{theorem:recursive}.

\begin{theo}\label{theorem:recursive:general}
Let $\mathcal{L}$ be as defined in~\eqref{Eq:indices:new} and let $d$ denote the $k$-th index in $\mathcal{L}$ such that $t_d=t_{\ell_k}$. The estimator~\eqref{Eq:batch:estimator} with the weighting matrix $W_{\ell,k}$ satisfying~\eqref{Eq:recursive:W:properties} may be computed recursively as follows
\begin{align}\label{Eq:recursive:estimator:d}
\hat{x}_k
&=P_k\big(W_{k-1,k}^2A_{k-1,k}^TP_{k-1}^{-1}\hat{x}_{k-1}
+W_{d,k}^2A_{d,k}^TH_d^TR_d^{-1}y_{d}\big),
\end{align}
where the estimator covariance, $P_k$, is given by
\begin{align}\non
P_k^{-1}
&=W_{k-1,k}^2A_{k-1,k}^TP_{k-1}^{-1}A_{k-1,k}\\\label{Eq:recursive:covariance:d}
&\qquad+W_{d,k}^2A_{d,k}^TH_d^TR_d^{-1}H_dA_{d,k}.
\end{align}
\end{theo}
It is important to emphasize the implications of Theorem~\ref{theorem:recursive:general}. The optimal update of the estimator is not affected by the times at which the measurements are acquired, as long as the actual validity time of these measurements is known. This means that seemingly different problems such as filtering and out-of-sequence measurement processing may be treated in essentially the same manner using~\eqref{Eq:recursive:estimator:d}. In other words, measurements arriving out-of-sequence are optimally processed using the same algorithm as those arriving in-sequence. This is not the case in the Bayesian setup, where a different algorithm is required to incorporate OOSM. Moreover, except for some special cases, standard OOSM processing is either suboptimal or requires increased computational resources. On the contrary, in the discussed setup, OOSM processing remains optimal in the sense of Theorem~\ref{theorem:recursive:general}.

Inspecting the filter equations of Theorem~\ref{theorem:recursive:general} we notice that additional generalization is in place. The first summand in both~\eqref{Eq:recursive:estimator:d} and~\eqref{Eq:recursive:covariance:d} stands for the contribution of the previously updated estimate to the current time. However, the time index $k$ does not have to refer to a specific time in the past and may, in fact, refer to any time in the past, present, or future. Therefore, we may rewrite the equations of the presented algorithm so that, in addition to the ``measurement update step'' defined by the second summand in~\eqref{Eq:recursive:estimator:d} and~\eqref{Eq:recursive:covariance:d} that allows arbitrary measurement timing, the ``time propagation step'' is also arbitrary as summarized in the following Theorem.
\begin{theo}\label{theorem:recursive:general:extra}
Let $\mathcal{L}$ be as defined in~\eqref{Eq:indices:new} and let $d$ denote the $k$-th index in $\mathcal{L}$ such that $t_d=t_{\ell_k}$. In addition, let $m$ refer to an arbitrary index representing time instance $t_m$ at which $\hat{x}_m$ and $P_m$ are known.
The estimator~\eqref{Eq:batch:estimator} with the weighting matrix~\eqref{Eq:recursive:W:properties} may be computed recursively as follows
\begin{align}\label{Eq:recursive:estimator:d:extra}
\hat{x}_k
&=P_k\big(W_{m,k}^2A_{m,k}^TP_{m}^{-1}\hat{x}_{m}
+W_{d,k}^2A_{d,k}^TH_d^TR_d^{-1}y_{d}\big),
\end{align}
where the estimator covariance, $P_k$, is given by
\begin{align}\label{Eq:recursive:covariance:d:extra}
P_k^{-1}
&=W_{m,k}^2A_{m,k}^TP_{m}^{-1}A_{m,k}+W_{d,k}^2A_{d,k}^TH_d^TR_d^{-1}H_dA_{d,k}.
\end{align}
\end{theo}
Clearly, Theorem~\ref{theorem:recursive:general:extra} generalizes both Theorem~\ref{theorem:recursive} and Theorem~\ref{theorem:recursive:general} which follow as special cases. At this point, we may further generalize the discussion that follows Theorem~\ref{theorem:recursive:general} by realizing that the problem of fixed-lag smoothing may be treated by the same algorithm as filtering and OOSM processing. While in the Bayesian setup this problem requires a modified algorithm (known as the fixed-lag smoother) and cannot be done using a standard Kalman filter, in our case the problems of filtering, fixed-lag smoothing and OOSM processing may be solved optimally using the same algorithm described in Theorem~\ref{theorem:recursive:general:extra}.

\section{Properties and Examples}\label{section:properties:examples}

\subsection{Examples}\label{section:examples1}
\subsubsection{Static Parameter}\label{section:examples1:static}
Let $A_{k,\ell}=I$ 
meaning that $x_k=x_{k-1}=\ldots=x_0$. In this case~\eqref{Eq:problem:WLS} reads
\begin{align}\label{Eq:example:WLS:static}
\arg\min_{x_k}\sum_{\ell\in\mathcal{L}}(H_\ell x_k-y_\ell)^T W_{\ell,k}R_\ell^{-1}W_{\ell,k}^T(H_\ell x_k-y_\ell).
\end{align}
We assign equal weights to all measurements in the set $\mathcal{L}$ by setting, for all $\ell,k$, $W_{\ell,k}=1$ and obtain the following form of the estimator~\eqref{Eq:batch:estimator}:
\begin{align}\label{Eq:example:estimator:constant}
\hat{x}_k&=P_k\sum_{\ell\in\mathcal{L}}H_\ell^TR_\ell^{-1}y_\ell,
\end{align}
where the covariance, $P_k$, is given by
\begin{align}\label{Eq:example:covariance:constant}
P_k&=\paren{\sum_{\ell\in\mathcal{L}}H_\ell^TR_\ell^{-1}H_\ell}^{-1}.
\end{align}
Note that $W_{\ell,k}=1$ satisfies the set of properties~\eqref{Eq:recursive:W:properties}. The corresponding recursive form of the estimator thus reads
\begin{align}\label{Eq:example:estimator:recursive:constant}
\hat{x}_k&=P_k(P_{k-1}^{-1}\hat{x}_{k-1}+H_k^TR_k^{-1}y_k)\\
P_k^{-1}&=P_{k-1}^{-1}+H_k^TR_k^{-1}H_k.
\end{align}

\subsubsection{Constant Acceleration Model}\label{section:examples1:polynomial}
Consider a state comprising position, velocity and acceleration such that
$
x_0
\triangleq\begin{pmatrix}{\rm pos}_0 & {\rm vel}_0 & {\rm acc}_0\end{pmatrix}^T
$.
The state follows the constant acceleration (CA) model with position-only measurements
\begin{align}\non 
x_k
&=
\begin{pmatrix}1 & \Delta t & \Delta t^2/2\\0 & 1 & \Delta t\\ 0 & 0 & 1\end{pmatrix}
\begin{pmatrix}{\rm pos}_{k-1}\\{\rm vel}_{k-1}\\{\rm acc}_{k-1}\end{pmatrix}
=
\begin{pmatrix}1 & t_k & t_k^2/2\\0 & 1 & t_k\\ 0 & 0 & 1\end{pmatrix}
x_0.
\end{align}
Here $\Delta t\triangleq t_k-t_{k-1}$ and $H_k=[1\ 0\ 0]$. Assuming the measurement noise variance is constant and using $W_{\ell,k}=1$ as before,~\eqref{Eq:batch:estimator} reduces to
\begin{align}\non
\hat{x}_k
&=
\begin{pmatrix}
\sum_\ell  t_\ell^0           & \sum_\ell t_\ell              & \frac{1}{2}\sum_\ell t_\ell^2\\
\sum_\ell t_\ell              & \sum_\ell t_\ell^2 & \frac{1}{2}\sum_\ell t_\ell^3\\
\frac{1}{2}\sum_\ell t_\ell^2 & \frac{1}{2}\sum_\ell t_\ell^3 & \frac{1}{4}\sum_\ell t_\ell^4
\end{pmatrix}^{-1}
\begin{pmatrix}
\sum_\ell t_\ell^0 y_\ell\\\sum_\ell t_\ell y_\ell\\\frac{1}{2}\sum_\ell t_\ell^2 y_\ell
\end{pmatrix},
\end{align}
which is the exact result of Savitzky-Golay~\cite{savitzky1964smoothing} for polynomial smoothing of noisy data. This observation allows us to relate the Savitzki-Golay filter to the more general KF framework. In other words, polynomial smoothing can be formulated within a state-space model using appropriate transition matrix~\eqref{Eq:problem:state}, and measurement geometry~\eqref{Eq:problem:measurement}.

\subsection{Properties}\label{section:properties}
We now state several properties satisfied by the batch and recursive versions of the estimator.
\begin{theo}\label{theorem:efficient}
The estimator~\eqref{Eq:batch:estimator} is unbiased and the corresponding error covariance~\eqref{Eq:batch:covariance} attains the Cram\'{e}r-Rao lower bound with equality rendering the estimator efficient.
\end{theo}

\begin{lemm}\label{lemma:scaling} The optimal estimator in either batch~\eqref{Eq:batch:estimator}, or recursive~\eqref{Eq:recursive:estimator} form does not depend on the scaling of the measurement noise covariance.
\end{lemm}
Lemma~\ref{lemma:scaling} states that knowing the measurement noise variance is only required up to a multiplicative constant. Thus, for a constant measurement noise variance, one does not need to tune the corresponding parameter in the filter. This desired property is only possible due to the lack of the process noise and does not happen in, e.g., Kalman-filter based algorithms.

\section{Exponentially Weighted Information Filtering}\label{section:wif}
\subsection{Concept}\label{section:wif:concept}
In this section we consider a specific function $W_{\ell,k}$ and discuss the resulting estimator. The function under consideration is~\eqref{Eq:recursive:W} which is repeated here for convenience.
\begin{align}\label{Eq:recursive:W2}
W_{\ell,k} = e^{-{(t_k-t_\ell)}/2\tau},\quad\tau>0.
\end{align}
It is readily seen that this weighting function satisfies the set of properties~\eqref{Eq:recursive:W:properties} such that a recursive form of the estimator~\eqref{Eq:batch:estimator} is given by~\eqref{Eq:recursive:estimator}. From this point on, however, we drop the formulation~\eqref{Eq:problem:measurement2} and return to the original setup~\eqref{Eq:problem:measurement}. Note that this does not affect the optimality of the solution in the weighted least-squares sense~\eqref{Eq:problem:WLS}. For this choice of $W_{\ell,k}$ the recursive estimator~\eqref{Eq:recursive:estimator}  takes the following form
\begin{align}\label{Eq:recursive:estimator:wif}
\hat{x}_k
&=P_k\big(e^{-\frac{t_k-t_{k-1}}{\tau}}A_{k-1,k}^TP_{k-1}^{-1}\hat{x}_{k-1}+H_k^TR_k^{-1}y_k\big),
\end{align}
where $P_k$ is given by
\begin{align}\label{Eq:recursive:covariance:wif}
P_k^{-1}=e^{-\frac{t_k-t_{k-1}}{\tau}}A_{k-1,k}^TP_{k-1}^{-1}A_{k-1,k}+H_k^TR_k^{-1}H_k.
\end{align}
Hereafter, we term the above estimator Exponentially Weighted Information Filter (EWIF). Note that once we have dropped the formulation~\eqref{Eq:problem:measurement2}, $P_k$ given in~\eqref{Eq:recursive:covariance:wif} is no longer the covariance of the estimator, but, rather, an approximation. This is similar to a KF setup in which the model is not known precisely and an artificial process noise is used to compensate this mismatch, such that the \emph{computed} covariance is not the true error covariance.


We note in passing that although exponential weighting has been addressed in the literature in the past~\cite{anderson1973exponential,anderson1979optimal}, this was never done as an approach to deterministically cope with model uncertainties. Thus, the setup considered in the present section is a novel approach to address state estimation problems under possible model mismatch.


\subsection{Relation to Kalman Filter}\label{section:wif:kalman}
In order to gain additional insight into the differences between the EWIF approach and a standard KF setup, we recall the standard KF equations.
\begin{align}\label{Eq:covform:kf:state}
\hat{x}_k
&=A_{k,k-1}\hat{x}_{k-1}+K_k(y_k-H_kA_{k,k-1}\hat{x}_{k-1}),
\end{align}
where $P_k$ is given by
\begin{align}\label{Eq:covform:kf:covariance}
P_k&=(I-K_kH_k)P_k^-,
\end{align}
the Kalman gain, $K_k$ reads
\begin{align}\label{Eq:covform:kf:gain}
K_k&=P_k^-H_k^T(H_kP_k^-H_k^T+R_k)^{-1},
\end{align}
and
\begin{align}\label{Eq:covform:kf:covpredict}
P_k^-=A_{k,k-1}P_{k-1}A_{k,k-1}^T+Q_{k,k-1}.
\end{align}
It will be convenient to consider the information form of the latter, which we express in a somewhat nonstandard form as follows
\begin{align}\non
\hat{x}_k
&=P_k\big(A_{k-1,k}^T(P_{k-1}+A_{k-1,k}Q_{k-1,k}A_{k-1,k}^T)^{-1}\hat{x}_{k-1}\\\label{Eq:information:kf:state}
&\qquad\qquad+H_k^TR_k^{-1}y_k\big),
\end{align}
where $P_k$ is given by
\begin{align}\non
P_k^{-1}
&=A_{k-1,k}^T(P_{k-1}+A_{k-1,k}Q_{k-1,k}A_{k-1,k}^T)^{-1}A_{k-1,k}\\\label{Eq:information:kf:covariance}
&\qquad\qquad+H_k^TR_k^{-1}H_k.
\end{align}
Showing the equivalence of the two formulations is a matter of straightforward algebraical manipulations. Comparing~\eqref{Eq:information:kf:state} and ~\eqref{Eq:information:kf:covariance} with, respectively,~\eqref{Eq:recursive:estimator:wif} and~\eqref{Eq:recursive:covariance:wif} we make the following observations.
It is readily seen that if $\tau\to\infty$ in~\eqref{Eq:recursive:estimator:wif} and~\eqref{Eq:recursive:covariance:wif}, $e^{-\frac{t_k-t_{k-1}}{\tau}}\to1$ and the dependence of $\hat{x}_k$ on the measurement history becomes uniform so that no extra weight is assigned to the more recent measurements. In this regime the algorithm reduces to a standard KF by setting $Q_{k-1,k}=0$ in~\eqref{Eq:information:kf:state} and ~\eqref{Eq:information:kf:covariance}. On the contrary, if $\tau\to0$, then $e^{-\frac{t_k-t_{k-1}}{\tau}}\to0$, and $\hat{x}_k$ depends solely on the current measurement $y_k$. This is similar to setting a very large process noise in the standard KF setup.

Further, in both cases, the algorithms linearly combine two types of data -- historical information, captured by the previously obtained estimate $\hat{x}_{k-1}$, and new information stored in the currently acquired measurement $y_k$. In the KF case, the weight assigned to the history is controlled by the quantity $(P_{k-1}+A_{k-1,k}Q_{k-1,k}A_{k-1,k}^T)^{-1}$ which is the information (inverse of the covariance) associated with older measurements. The reduction in the information, relatively to $P_{k-1}^{-1}$ from step $k-1$, is dictated by the process noise covariance $Q_{k-1,k}$ such that in the case of a very unpredictable state sequence (large process noise), much uncertainty is added between two consecutive measurement updates and the amount of information stored in the previous measurements reduces. On the contrary, at the absence of process noise, the weight assigned to the history of the measurements in the EWIF approach is controlled by the quantity $e^{-\frac{t_k-t_{k-1}}{\tau}}P_{k-1}^{-1}$. Since for $t_k\geq t_{k-1},\tau\geq0$,
\[
e^{-\frac{t_k-t_{k-1}}{\tau}}\leq1,
\]
the information associated with the measurement history reduces similarly to the KF case. In other words, the difference between the KF-based approach and the proposed EWIF method boils down to how the contribution of historical measurements is managed -- additive increase of the error covariance in the case of KF versus multiplicative inflation of the covariance in EWIF.

Our final comment relates to specific information reduction mechanism. While in the standard KF setup the weight assigned to the measurement history is controlled by the state process noise or, equivalently, by our assumptions on the system predictability, in the proposed approach this weight control mechanism is completely decoupled from the state and only considers the measurement sequence. This exposes additional robustness of EWIF to system model mismatch, similarly to the Savitzky-Golay filter. The parameter $\tau$ becomes then nothing but a tuning parameter as opposed to the Bayesian case where the process noise covariance has to be carefully chosen for each state dynamics based on some prior knowledge.


\subsection{Toy Example}\label{section:wif:toy}
\begin{figure}[h]\centering
{\includegraphics[width=\columnwidth, trim=0cm 0.1cm 0cm 0.1cm, clip=true]{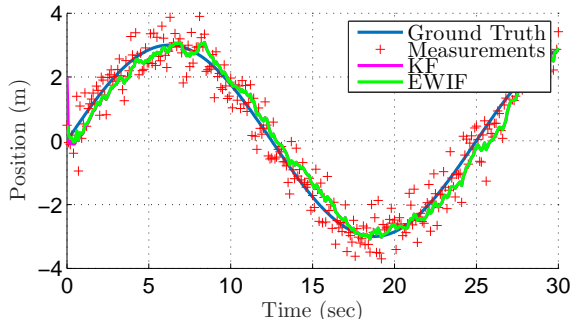}}
\caption{Tracking a sine signal using EWIF (solid green) and KF (solid magenta). The two estimated sequences are practically indistinguishable -- the RMS errors of EWIF and KF are $0.31$ and $0.33$, respectively.}
\label{fig:toy:sine}
\end{figure}

We now demonstrate the utility of the EWIF approach in a simple synthetic example. We consider a one-dimensional sinusoidal signal observed by a sequence of linear measurements corrupted by zero-mean Gaussian noise with variance $0.5$.
Note that the state at $k+1$ cannot be described as a linear transformation of the state at $k$. Thus, in order to track the sequence with a \emph{linear} filter one has to compensate the model mismatch. We estimate the sequence of states using a linear Kalman filter in which the model mismatch is compensated by a process noise such that the state equation is
\[
x_k=x_{k-1}+w_k,
\]
where $\set{w_k}$ is a zero-mean white process noise sequence with standard deviation $\sigma_w=0.1$. Alternatively, the states are tracked using the EWIF algorithm with decorrelation constant $\tau=0.5$. The resulting estimates are presented in Fig.~\ref{fig:toy:sine}.
It is readily seen that, for the described parameter setup, both algorithms achieve similar performance. This demonstrates the history handling mechanism carried out by the two methods as discussed in Section~\ref{section:wif:kalman} and also suggests that, in similar estimation problems, for a given value of $\sigma_w$, one can find an ``equivalent'' value of $\tau$.



\section{Applications}\label{section:applications}
In this section we consider several classical applications commonly addressed and show how these can be treated with the considered deterministic framework.
\subsection{Prediction and Smoothing}\label{section:prediction}
In order to derive optimal, in the sense of Section~\ref{section:derivation}, predictor we consider~\eqref{Eq:recursive:estimator:d:extra} and~\eqref{Eq:recursive:covariance:d:extra} for some index that $m$ representing $t_m$ at which $\hat{x}_m$ and $P_m$ are known and $k$ is an index representing some future time $t_k\triangleq t_m+\Delta$. It is assumed that no additional informative measurements are available for the computation of $\hat{x}_k$. This is equivalent to setting infinite covariance to any new measurement thus nullifying the second summand in~\eqref{Eq:recursive:estimator:d:extra} and~\eqref{Eq:recursive:covariance:d:extra} which then become
\begin{align}\label{Eq:recursive:estimator:prediction}
\hat{x}_k
&=A_{k,m}\hat{x}_{m}\\\label{Eq:recursive:covariance:prediction}
P_k&=\tfrac{1}{W_{m,k}^2}A_{k,m}P_{m}A_{k,m}^T.
\end{align}
Note that the state prediction equation is exactly the same as in the KF case and the difference in the covariance expression is along the lines of the discussion in Section~\ref{section:wif:kalman}.
Some smoothing problems are trivially obtained from~\eqref{Eq:recursive:estimator:d:extra} and~\eqref{Eq:recursive:covariance:d:extra}. For example, 
fixed-lag smoothing is obtained similarly by considering a constant lag ${t_k-t_m}$ in the above equations.
\subsection{Out-of-Sequence Measurement Processing}\label{section:oosm}
In centralized multisensor fusion systems
it is common practice that some measurements arrive ``out-of-sequence'', namely, after other measurements, with later time-tags, have already been processed. 
The main challenge in utilizing OOSM, in the Bayesian setup, is the correlation generated by the past measurements and the process noise sequence and significant effort has been done in this direction. Some recent contributions, in addition to those cited in Section~\ref{section:intro}, include~\cite{koch:2011:oosm:accumulated},~\cite{zhang2012optimal} and references therein. In all cases, a significant modification of the KF is required in order to cope with OOSM either exactly or approximately. In~\cite{koch:2011:oosm:accumulated}, e.g., state augmentation is required, thus increasing the computation burden when utilizing OOSM.

Next, we demonstrate how the OOSM utilization may be done in a simple manner using the EWIF approach.
First, we rewrite~\eqref{Eq:recursive:estimator:d} and~\eqref{Eq:recursive:covariance:d} using the weighting function~\eqref{Eq:recursive:W2}.
\begin{align}\non
\hat{x}_k
&=P_k\big(e^{-\frac{t_k-t_{k-1}}{\tau}}A_{k-1,k}^TP_{k-1}^{-1}\hat{x}_{k-1}\\\label{Eq:recursive:estimator:d:oosm}
&\qquad\qquad\qquad+e^{-\frac{t_k-t_{d}}{\tau}}A_{d,k}^TH_d^TR_d^{-1}y_{d}\big),
\end{align}
where the estimator covariance, $P_k$, is given by
\begin{align}\non
P_k^{-1}
&=e^{-\frac{t_k-t_{k-1}}{\tau}}A_{k-1,k}^TP_{k-1}^{-1}A_{k-1,k}\\\label{Eq:recursive:covariance:d:oosm}
&\qquad\qquad\qquad+e^{-\frac{t_k-t_{d}}{\tau}}A_{d,k}^TH_d^TR_d^{-1}H_dA_{d,k}.
\end{align}
Recall that $t_k$ is the time at which the state update is required and the current measurement is acquired, $t_{k-1}$ is the previous update time, and $t_d=t_{\ell_k}$ is the time for which the currently acquired measurement, $y_{\ell_k}\equiv y_d$, is valid. If $y_{\ell_k}$ is an in-sequence measurement, i.e., $t_d=t_k$, then $e^{-\frac{t_k-t_{d}}{\tau}}=1$ and the update rule reduces to that of~\eqref{Eq:recursive:estimator}. If $t_d\neq t_k$, the contribution of $y_{\ell_k}$ is reduced by an exponential factor accounting for the temporal distance of $y_{\ell_k}$ from the current time. It follows that processing an OOSM differs from utilizing an in-sequence measurement by the computation of an exponent of a scalar and, practically, requires no extra computational effort. This does not occur in \emph{any} Bayesian approach.

One may wonder whether the performance of the algorithm depends on the order the measurements arrive. Invariance to the order of data processing is a desirable property of an algorithm. It is clear that, in the Bayesian setup, MMSE-optimal processing (which is not always feasible) is invariant to the order data are processed. Observing~\eqref{Eq:recursive:estimator:d:oosm} and~\eqref{Eq:recursive:covariance:d:oosm} it becomes evident that the same occurs in the proposed approach. Indeed, both of the above equations depend solely on the measurement validity time and not on the time these measurements arrived. Another way to arrive at the same conclusion is by observing the batch form of the estimator~\eqref{Eq:batch:estimator}. It is readily seen that different orders of data processing correspond to different orders the summations in~\eqref{Eq:batch:estimator} are performed. Clearly, all yield identical results.

\begin{figure}[h!]\centering
{\includegraphics[width=1.1\columnwidth, trim=2.5cm 0cm -0.5cm 0.5cm, clip=true]{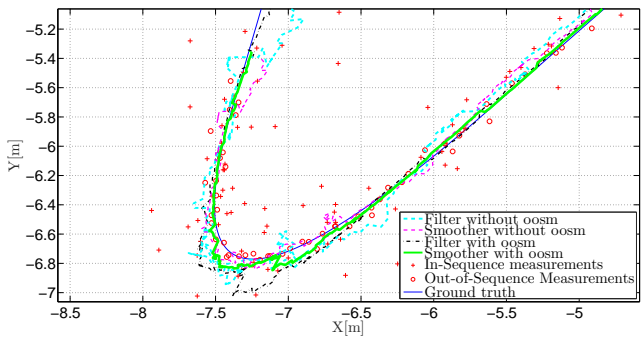}}
\caption{Tracking a target with arbitrary dynamics under four measurement regimes using the same EWIF algorithm with $\tau=2$ (sec). A two-dimensional CA model is used as the target model. The blue thin line is the true trajectory. Pluses and circles are the in-sequence and out-of-sequence measurements, respectively. Dashed cyan and magenta curves are, respectively, the filter and $1$-sec-lag smoother using only the in-sequence measurements. Black and green curves are, respectively, the filter and $1$-sec-lag smoother using the in-sequence measurements and OOSM. An animation of the complete scenario can be found at \\ \protect\url{https://youtu.be/2G1nEoNwwjY}.}
\label{fig:movie:sample}
\end{figure}

In order to demonstrate the algorithm for OOSM, we consider an example in which target moves arbitrarily in 2D space. Its position is observed by two sensors, one of which is undergoing a delay. The noise variance and measurement frequency vary along the scenario. The delay durations change randomly. The same EWIF algorithm summarized in~\eqref{Eq:recursive:estimator:d:oosm} and~\eqref{Eq:recursive:covariance:d:oosm} is used to track the target under four measurement regimes -- filtering and smoothing, with and without incorporating OOSM. A snapshot of the trajectory accompanied by the corresponding measurements and estimates is presented in Fig.~\ref{fig:movie:sample}. An animation of the complete scenario can be found at \protect\url{https://youtu.be/2G1nEoNwwjY}.

A sample implementation of the EWIF algorithm for a simpler example of fixed-lag smoothing in the presence of OOSM may be found at\\
\protect{\url{https://tinyurl.com/y3e4df3s}}.



\section{Concluding Remarks}\label{section:concluding}
We presented a deterministic framework for state estimation in dynamical systems. As opposed to the classical approach, where (random) process noise is used to cope with inevitable uncertainties in the dynamical model, in the presented setup, this uncertainty is treated by assigning exponentially decaying weights to the observations, so that more recent measurements are assigned more weight than ``older'' ones. The state is then estimated using a WLS algorithm. For the chosen weighting function, the algorithm possesses a convenient recursive form similarly to the Kalman filter. Handling the inherent uncertainty in the system model by using the weighted information approach allowed us to treat several, seemingly different problems using essentially the same algorithm. These include filtering, fixed-lag smoothing and OOSM processing. Lastly, we refer to our first example to emphasize that it is not a-priori obvious whether the proposed approach is preferable over the classical usage of process noise and a definite answer is, probably, case-dependent. A clear advantage of the present approach is its applicability to a variety of estimation problems resulting in the same algorithm.



%% file: intro_mfi.tex
In the standard setup of the Kalman filter (KF) an unknown state sequence is modeled as a stochastic process evolving through a linear system and driven by an external random process noise. The sequence is observed by a linear measurement channel contaminated by random measurement noise. The resulting estimator is an optimal, in the minimum mean-square error (MMSE) sense, recursive algorithm for the estimation of the state using the available measurements.

The process noise sequence may be interpreted in two ways. From a mathematical point of view, it is the actual generator of the state sequence. In this case, the state and measurement models are assumed to be known perfectly and state estimation becomes a mathematical problem of Bayesian inference of a hidden sequence from noisy measurements. From a practical perspective, on the other hand, process noise is the designer's tool to cope with
system uncertainties. For example, a (possibly) nonlinear dynamics governing the actual system's behavior may be modeled using a simple, linear motion model with process noise having sufficiently high variance. Clearly, the closer the system model to the true dynamics, the better the filter performance is expected to be. However, since system uncertainties are inevitable, using process noise cannot be avoided. The process noise is, thus, a method for handling unmodeled uncertainties in the dynamical (and, in fact, measurement) models.

Numerous generalizations and modifications to the basic version of the KF have been proposed. Optimal prediction and smoothing algorithms~\cite{anderson1979optimal} are only two examples. A more recent notion of out-of-sequence measurement (OOSM) processing has become popular since the publication~\cite{ybs:2002:oosm}. In this setup, some measurements are delayed and arrive after other, more recent measurements have already been processed. It is then required to update state estimates using the older measurements without reprocessing the entire measurement sequence. In some cases, such as Kalman smoothing and one-step OOSM processing~\cite{ybs:2002:oosm}, the derived algorithms remain optimal. There are scenarios, however, in which preserving optimality without increasing computational requirements is no longer possible. The problem of multi-step OOSM processing is a good example of such a scenario. Typically, the difficulty in deriving an algorithm or maintaining its optimality are rooted in the correlations generated by the process noise sequence and earlier measurements. On the other hand, the process noise is an inevitable mechanism in order to treat mismatches in the dynamical model of the state sequence. Consequently, some existing solutions to the multistep OOSM processing problem, e.g.,~\cite{ybs:2004:oosm:multistep,zhang2010tracking}, make approximations thus sacrificing optimality. Other methods that address this problem increase the computation burden by augmenting the state vector (see~\cite{koch:2011:oosm:accumulated,govaers:2014:oosm:generalized}).

Despite their similar form, the resulting optimal (or suboptimal) algorithms are different in all of the aforementioned cases. For example, the gain matrix computed for the fixed-lag smoothing problem is computed differently than the gain for utilizing one-step OOSM. In other words, it is required to use a unique algorithm for each of the above problems.

In this paper we propose a different mechanism for handling unmodeled system uncertainties. Instead of injecting random process noise, we assign different weights to different measurements so that more recent measurements are assigned more weight. In this setup, the state at any time is a deterministic function of the state at any other time. We formulate a weighted least-squares cost for the estimation of the state using the available measurements and derive the corresponding solution. In this regard, the proposed approach is a generalization of the Savitzky-Golay filter~\cite{savitzky1964smoothing} for polynomial smoothing of noisy data. A specific choice of an exponentially decaying weight function results in an algorithm with essentially the same recursive structure as the Kalman filter. An important difference, however, is the manner in which old data are combined with newly obtained measurements. While in the classical KF, the uncertainty associated with the previous estimate is inflated in an additive manner (by adding the process noisy covariance to the state error covariance matrix), in our case, the uncertainty inflation is done by multiplying the previous covariance matrix by an exponential factor. This simple difference allows us to solve a larger variety of problems using essentially the same algorithm. These problems include filtering, prediction, smoothing and general out-of-sequence updates, all of which, as mentioned, require a unique Kalman-like algorithm when addressed in a classical manner. In addition, the proposed algorithm does not require additional computational effort when applied to more complex problems.

The remainder of the paper is organized as follows. In Section~\ref{section:problem} we formally state the problem at hand and list the relevant assumptions. Section~\ref{section:derivation} is devoted to the derivation of several algorithms starting with the simplest case of a batch-type estimator and proceeding to more complex, recursive routines. In Section~\ref{section:properties:examples} several important properties of the derived estimator are discussed and some illustrative simple examples are provided. In Section~\ref{section:wif} we consider an important special case of an exponentially decaying weight function and compare the resulting algorithm with the classical KF. In Section~\ref{section:applications} several applications are considered, the main of which is OOSM processing in the filtering and smoothing framework. Concluding remarks are given in Section~\ref{section:concluding}. Due to space limitations, the proofs of the theorems are omitted.

%% file: problem_mfi.tex
We consider a dynamical system evolving in an arbitrary manner in time. The evolution of the system is \emph{modeled} by the following \emph{deterministic} equation
\begin{align}\label{Eq:problem:state}
x_k = A_{k,\ell}x_\ell+C_{k,\ell}u_{k,\ell},
\end{align}
where $x_k$ is the system state at time $t_k$, $A_{k,\ell}$ is the known state transition matrix from time $t_\ell$ to time $t_k$ and $u_{k,\ell}$ is the cumulative effect of the deterministic inputs applied during the time interval $(t_\ell,t_k]$. The initial state $x_0$ is assumed to be deterministic but unknown. Clearly, the simple modeling may deviate significantly from the true system dynamics. In this case, there will be a model mismatch. We emphasize that we do not assume that the system is driven by a (random) process noise meaning that the only uncertainty in the described system is the initial state $x_0$. In other words, no probabilistic modeling is associated with the sequence $\set{x_k}$. For all $k,\ell\in\mathbb{N}$, the matrices $\set{A_{k,\ell}}$  are assumed to be invertible~\cite{ybs:2002:oosm}. Similarly to~\eqref{Eq:problem:state}, assuming appropriate controllability conditions, we may express $x_\ell$ using $x_k$ as follows
\begin{align}\label{Eq:problem:state:reverse}
x_\ell = A_{\ell,k}x_k+C_{\ell,k}u_{\ell,k},
\end{align}
where $A_{\ell,k}=A_{k,\ell}^{-1}$ and $C_{\ell,k}u_{\ell,k}=-A_{k,\ell}^{-1}C_{k,\ell}u_{k,\ell}$. This allows the formulation of a direct observation of $x_\ell$ as follows
\begin{align}\label{Eq:problem:measurement}
y_\ell = H_\ell x_\ell+v_{\ell}.
\end{align}
Here $H_\ell$ is a known matrix, and $\set{v_\ell}$ is a sequence of independent Gaussian random variables such that $v_\ell\sim\gaus{0}{R_\ell}$. Note that, using~\eqref{Eq:problem:state:reverse},~\eqref{Eq:problem:measurement} also serves as an indirect observation of $x_k$.

Our goal is estimating the state $x_k$ using the available measurements $\set{y_\ell,\,\ell\in\mathcal{L}}$  for some set of indices $\mathcal{L}$. Note that since the state sequence is deterministic, every measurement in the set $\mathcal{L}$ may be viewed as a measurement of $x_k$ without adding uncertainty (in addition to the measurement noise $v_\ell$). This is different from the Bayesian setup in which, in addition to the measurement noise, there is additional uncertainty due to the contribution of the process noise in the time interval defined by the considered measurement and the desired state $x_k$. In the absence of a probabilistic prior for $x_k$, and taking into account the symmetry of the distribution of $v_k$, we utilize the relation~\eqref{Eq:problem:state:reverse} and consider the following weighted least-squares (WLS) cost
\begin{align}\nonumber
&\operatorname{argmin}_{x_k}\sum_{\ell\in\mathcal{L}}(H_\ell(A_{\ell,k}x_k+C_{\ell,k}u_{\ell,k})-y_\ell)^T\\\label{Eq:problem:WLS}
&\qquad \times W_{\ell,k}R_\ell^{-1}W_{\ell,k}^T(H_\ell(A_{\ell,k}x_k+C_{\ell,k}u_{\ell,k})-y_\ell).
\end{align}
In other words,~\eqref{Eq:problem:WLS} is a sum of normalized squared differences between the measurements $y_\ell$ and linear predictions thereof given by $H_\ell(A_{\ell,k}x_k+C_{\ell,k}u_{\ell,k})\equiv H_\ell x_\ell$. The weighting matrix of the $\ell$-th element in the above equation comprises two elements -- the inverse covariance matrix of the corresponding measurement noise and an additional matrix $W_{\ell,k}$ which will play a crucial role in the sequel. Intuitively, the dependence of $W_{\ell,k}$ on $k$ causes this matrix to account for the relevance of the distant measurement $y_\ell$ to the estimation of the present state $x_k$ thus controlling the contribution of $y_\ell$ in case of a model mismatch. Perhaps the simplest example of $W_{\ell,k}$ is the following, scalar-valued
Savitzky-Golay~\cite{savitzky1964smoothing} kernel
\begin{align}\label{Eq:problem:sg:kernel}
W_{\ell,k}=
\begin{cases}
1,& \size{t_\ell-t_k}\leq1\\
0,&\rm{otherwise}
\end{cases}.
\end{align}
Additional discussion on $W_{\ell,k}$ and its properties is presented in Sections~\ref{section:derivation} and~\ref{section:wif}.

Note that the cost~\eqref{Eq:problem:WLS} could be obtained for the following formulation of the measurement equation for some index $k$,
\begin{align}\label{Eq:problem:measurement2}
y_\ell = H_\ell x_\ell+\tilde{W}_{\ell,k}v_{\ell},
\end{align}
where $\tilde{W}_{\ell,k}={W}_{\ell,k}^{-T}$ and the effective covariance is
\begin{align}\label{Eq:problem:measurement2:covariance}
\EE{\tilde{W}_{\ell,k}v_{\ell}v_\ell^T\tilde{W}_{\ell,k}^T}
    =\tilde{W}_{\ell,k}R_\ell\tilde{W}_{\ell,k}^T
    ={W}_{\ell,k}^{-T}R_\ell{W}_{\ell,k}^{-1}.
\end{align}
We note in passing that, although defining different from~\eqref{Eq:problem:measurement} measurements, this model yields a standard WLS cost with optimal weights meaning its solution is the best linear unbiased estimator.

In the following section we derive optimal, in the sense of~\eqref{Eq:problem:WLS}, estimator and consider special cases in which recursive formulation of the estimator is possible.